\documentclass[prb,aps,twocolumn,showpacs]{revtex4}
\usepackage{amssymb}
\usepackage{graphicx}

\begin{document}

\title{Self-Consistent Field Theory\\
of Brushes of Neutral Water-Soluble Polymers}
\author{Vladimir A. Baulin, Ekaterina B. Zhulina,$^\dag$ Avi Halperin}
\affiliation{Service des Interfaces et des Mat\'{e}riaux Mol\'{e}culaires et Macromol\'{e}%
culaires, DRFMC, CEA-Grenoble, 17 rue des Martyrs, 38054 Grenoble,
Cedex 9, France \vskip 0.5cm $^\dag$permanent address: Institute
of Macromolecular Components of the Russian Academy of Sciences,
199004 St Petersburg, Russia}


\begin{abstract}
The Self-Consistent Field theory of brushes of neutral water-soluble
polymers described by two-state models is formulated in terms of the
effective Flory interaction parameter $\chi _{eff}(T,\phi)$ that depends on
both temperature, $T$ and the monomer volume fraction, $\phi$. The
concentration profiles, distribution of free ends and compression force
profiles are obtained in the presence and in the absence of a vertical phase
separation. A vertical phase separation within the layer leads to a
distinctive compression force profile and a minimum in the plot of the
moments of the concentration profile vs. the grafting density. The analysis
is applied explicitly to the Karalstrom model. The relevance to brushes of
Poly(N-isopropylacrylamide)(PNIPAM) is discussed.

{\small \bf \center Accepted for publication in the Journal of
Chemical Physics}
\end{abstract}

\maketitle

\section{Introduction\label{introduction}}

\begin{figure}
\begin{center}
\includegraphics[width=6cm]{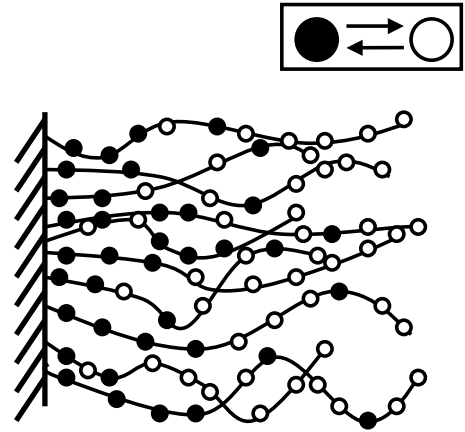}
\end{center}
\caption{Schematic picture of a brush of a "two-state polymer".
Open and filled circles depict monomers in the different,
interconverting, monomeric states.} \label{fig1}
\end{figure}

A number of "two-state" models were proposed to rationalize the phase
behavior of Poly(ethylene oxide) (PEO) and its solution thermodynamics.\cite%
{Karlstrom,Matsuyama,n-Cluster1,n-Cluster2,Bekiranov,DormidPhase} Within
these models the monomers are in dynamic equilibrium involving two
interconverting states (Fig. \ref{fig1}). The Flory-Huggins lattice and the
mixing entropy of the chains are retained. Additional contributions are due
to the mixing entropy of the different monomeric states and their
interactions. While the two-state models were proposed for aqueous solutions
of PEO they are also candidates for the description of other neutral
water-soluble polymers such as Poly(N-vinylpyrrolidone) (PVP) and
Poly(N-isopropylacrylamide) (PNIPAM).\cite{water2} The \textit{equilibrium }%
free energy obtained from these models can be expressed as\cite{BH} $%
F/kT=N^{-1}\phi \ln \phi +(1-\phi )\ln (1-\phi )+\chi _{eff}\phi (1-\phi )$
where $N$ is the polymerization degree. In distinction to the familiar Flory
free energy, the effective Flory interaction parameter $\chi _{eff}$ is a
function of both the monomer volume fraction, $\phi $, and the temperature, $%
T$. In this formulation the specific features of a particular
model are grouped into $\chi _{eff}(T,\phi )$. In the following we
consider the Self-Consistent Field (SCF) theory of brushes of
"two-state polymers" in terms of $\chi _{eff}(T,\phi )$. A
significant part of our discussion is devoted to brushes of
polymers capable of undergoing a second type of phase
separation.\cite{footXX, n-Cluster1, n-Cluster2, Solc1, Solc2}
Within a brush, this type of phase separation can lead to a
vertical phase separation associated with a discontinuous
concentration profile.\cite{Wagner,BH2} Our analysis focuses on
the signatures of such phase separation. These include the
non-monotonous variation of the brush thickness with the grafting
density and the appearance of distinct regimes in the compression
force profiles.

This approach is of interest because of a number of reasons: (i) $\chi
_{eff}(T,\phi )$ determines a number of important characteristics of the
brush among them the concentration profile, the distribution of free-ends
and the force profile associated with the compression of the brush. Thus, a
description of the brush behavior in terms of $\chi _{eff}(T,\phi )$
accounts for the leading brush properties and facilitates the comparison of
the predictions of the different models. The specific features of the
individual models and their parameters come into play when the distribution
of the monomer states is of interest. However, as we shall discuss, even in
this case it is convenient to first specify the brush characteristics in
terms of $\chi _{eff}(T,\phi )$. (ii) The formulation of the theory in terms
of $\chi _{eff}(T,\phi )$ underlines the relationship to the measurable
\[
\overline{\chi }(T,\phi )=\chi _{eff}-(1-\phi )\partial \chi _{eff}/\partial
\phi
\]%
as obtained from the study of the colligative properties of the
polymer solutions.\cite{Flory,wolfBook,Flory1,rev} $\overline{\chi
}(T,\phi )$ is helpful in determining the parameters of the
models. In the context of brushes, the behavior of $\overline{\chi
}(T,\phi )$ provides, as we shall discuss, a useful diagnostic for
systems expected to exhibit a vertical phase separation within the
brush. (iii) While our discussion focuses on the two-state models
cited above, the analysis can be extended to other
models\cite{sanchez,Freed} that yield a $\phi $ dependent $\chi
_{eff}$. (iv) The SCF theory of brushes characterized by $\chi
_{eff}(T,\phi )$ suggests useful tests for the occurrence of
vertical phase separation. This is of interest, as we shall
discuss in the final section, because of experimental indications
that such behavior occurs in brushes of PNIPAM. (v) The
concentration profiles obtained from the SCF theory are
essentially identical to those derived\cite{BH2} from the Pincus
approximation where the distribution of free-ends is assumed
rather than derived.\cite{Pincus,Pincus1} In marked contrast, the
compression force profiles are sensitive to the distribution of
free-ends and the two methods yield different results.

The two-state models differ in their identification of the
interconverting states. Within the $n$-cluster
model\cite{n-Cluster1,n-Cluster2} one is a bare monomer while the
second is a monomer incorporated into a stable cluster of $n$
monomers. In the remaining models one of the monomeric states is
hydrophilic and the other is hydrophobic. The hydrophilic state is
preferred at low $T$ while the hydrophobic state is favored at
high $T$. In the Karlstrom model\cite{Karlstrom} the two states
differ in their dipole moment and their interconversion involves
an internal rotation. The models of Matsuyama and
Tanaka,\cite{Matsuyama} Bekiranov \textit{et al}\cite{Bekiranov}
and of Dormindotova\cite{DormidPhase} assume that the hydrophilic
monomeric state forms a H-bond to a water molecule while the
hydrophobic state does not. The brush structure within the
Karlstrom model was studied using numerical SCF theory of the
Scheutjens-Fleer type\cite{Linse,Bjorling} and allowed to
rationalize the aggregation behavior of copolymers incorporating
PEO blocks.\cite{Linse1} The brush structure within the
$n$-cluster model was studied using SCF theory\cite{Wagner} and by
simulations.\cite{Mattice} These reveal the possibility of a
vertical phase separation within the brush giving rise to a
discontinuity in the concentration profile. In turn, this was
invoked in
order to rationalize observations about the collapse of PNIPAM brushes.\cite%
{Napper} The force profiles due to the compression of brushes described by
the $n$-cluster model were also analyzed.\cite{AHcluster,Sevick} The studies
of brushes of ''two-state polymers'' focused on a particular model and were
formulated in terms of the corresponding free energy. This obscured common
features between the different models and hampered the comparison between
them. For example, while a vertical phase separation is possible within all
two-state models, this scenario was mainly studied for the $n$-cluster model
thus creating a misleading impression about the physical origins of this
phenomenon.

Our discussion concerns a brush of flexible \textquotedblright
two-state\textquotedblright\ chains, terminally grafted to a planar surface.
We assume that the chains are monodisperse and that each chain incorporates $%
N$ monomers. The surface area per chain, $\sigma $, is constant and the
surface is assumed to be non-adsorbing for the two monomeric states. The
free energy per lattice site is
\begin{equation}
f_{\infty }(\phi ,T)/kT=(1-\phi )\ln (1-\phi )+\chi _{eff}(\phi ,T)\phi
(1-\phi )  \label{Ffint}
\end{equation}%
This form corresponds to the $N\rightarrow \infty $ limit. It is appropriate
for brushes of any $N$ because the grafted chains lose their mobility and
thus have no translational entropy. The application of our analysis to a
particular case is illustrated for the Karlstrom model. However, most of our
analysis is model independent in that $\chi _{eff}(T,\phi )$ is not
specified explicitly. The only assumption made is that $\overline{\chi }%
(T,\phi )$ can be expanded in powers of $\phi $
\[
\overline{\chi }(T,\phi )=\sum_{i=0}\overline{\chi }_{i}(T)\phi ^{i}
\]%
where the $\overline{\chi }_{i}(T)$ are specific to a given model. For
simplicity we further limit the discussion to systems where the first three
terms provide an accurate description of $\overline{\chi }(T,\phi )$. As we
shall discuss, the presence of a third order term is the minimal condition
for the possibility of a vertical phase separation within the brush. The
power series expansion of $\overline{\chi }(T,\phi )$ is clearly related to
the virial expansion of the osmotic pressure, $\pi $. The two differ in that
the second incorporates terms originating in the translational entropy. In
discussions of the SCF theory $f_{\infty }(\phi )/kT$ was often approximated
by the second and third terms in the virial expansion of the Flory-Huggins
free energy. From this point of view it is important to note two points: (i)
in the power series of $\overline{\chi }(T,\phi )$ all the coefficients are $%
T$ dependent and can change sign. (ii) Use of $\overline{\chi }(T,\phi )$
series expansion including a $\phi ^{3}$ term corresponds to a virial
expansion incorporating a $\phi ^{4}$ term.

The next five sections, \ref{SCFtheory}--\ref{distributionKarlsrom}, are
devoted to the model independent aspects of the SCF theory based on $%
f_{\infty }(\phi ,T)/kT$ with $\phi $ dependent $\chi _{eff}$. In section %
\ref{SCFtheory} we formulate the analytical SCF model for $\chi _{eff}(\phi
,T)$ as a generalization of the familiar SCF theory of brushes.\cite%
{Semenov,Milner,MilnerBrush,Skvortsov,Zhulina} The concentration profiles
and their moments are discussed in section \ref{concentrationProf}. The
technical details corresponding to section \ref{concentrationProf} are
described in Appendix \ref{appendixCalc}. Section \ref{distribution}
discusses the distribution of free ends while the technical details are
given in Appendix \ref{appendixg}. The force profiles associated with the
compression of the brush are analyzed in section \ref{compression}. In every
case we distinguish between brushes characterized by a continuos
concentration profile and those exhibiting a vertical phase separation.
Finally, in section \ref{distributionKarlsrom} we illustrate the
implementation of our analysis to the Karlstrom model. In particular, we
obtain the corresponding $\chi _{eff}(T,\phi )$, $\overline{\chi }(T,\phi )$
and the distribution of monomeric states in the brush.

\section{The SCF theory for $\protect\chi _{eff}(T,\protect\phi )$\label%
{SCFtheory}}

Consider a brush of neutral and flexible polymers comprising $N$
monomers of size $a$. Each chain is grafted by one end onto an
impermeable, non-adsorbing, planar surface. The area per chain is
denoted by $\sigma $ and $H$ is the maximal height of the brush.
Following refs. \onlinecite{Skvortsov,Zhulina}, the free energy
per chain, $F_{chain}$, consists of two terms: an interaction free
energy, $F_{int}$, and an elastic free energy, $F_{el}$,
$F_{chain}=F_{int}+$ $F_{el}$. The interaction free energy per
chain is

\begin{equation}
\frac{F_{int}}{kT}=\frac{\sigma }{a^{3}}\int_{0}^{H}f_{\infty }(\phi )dz
\label{Fint}
\end{equation}%
where the interaction free energy density $f_{\infty }(\phi )$ is given by (%
\ref{Ffint}). In a strong stretching limit, when the chains are extended
significantly with respect to Gaussian dimensions, the elastic free energy is%
$^{\mathbf{27}}$

\begin{equation}
\frac{F_{el}}{kT}=\frac{3}{2a^{2}}\int_{0}^{H}g(z^{\prime })dz^{\prime
}\int_{0}^{z^{\prime }}E(z,z^{\prime })dz.  \label{Fel}
\end{equation}
Here $E(z,z^{\prime })=dz/dn$ characterizes the local chain stretching at
height $z$ when the free end is at height $z^{\prime }$. $g(z^{\prime })$
specifies the height distribution of the free ends and obeys the
normalization condition $\int_{0}^{H}g(z^{\prime })dz^{\prime }=1$.

The concentration of monomers, $\phi (z)$, at height $z$ is specified by

\begin{equation}
\phi (z)=\frac{a^{3}}{\sigma }\int_{z}^{H}\frac{g(z^{\prime })dz^{\prime }}{%
E(z,z^{\prime })}  \label{ConFi}
\end{equation}
Since each chain consists of $N$ monomers we have

\begin{equation}
N=\frac{\sigma }{a^{3}}\int_{0}^{H}\phi (z)dz  \label{ConfN1}
\end{equation}
At the same time

\begin{equation}
N=\int_{0}^{z^{\prime }}\frac{dz}{E(z,z^{\prime })}  \label{ConN}
\end{equation}
which can be regarded as a normalization condition for the function $%
E(z,z^{\prime })$.

The equilibrium $\phi (z)$ in the brush is determined by the variation of
the functional $F_{chain}$ with respect to $E(z,z^{\prime })$ and $%
g(z^{\prime })$ subject to the constraints (\ref{ConfN1}) and (\ref{ConN})
yielding

\begin{equation}
E(z,z^{\prime })=\frac{\pi }{2N}\sqrt{z^{\prime 2}-z^{2}}  \label{E}
\end{equation}
and

\begin{equation}
\mu (\phi )=\lambda -Bz^{2}  \label{profile}
\end{equation}
where $B=3\pi ^{2}/8N^{2}a^{2}$, $\lambda $ is the Lagrange multiplier
associated with constraint (\ref{ConfN1}) and $\mu (\phi )=\partial
f_{\infty }(\phi )/\partial \phi $ is the exchange chemical potential. Up to
this point the SCF theory is identical to the familiar versions, as obtained
for $\chi =\chi (T)$.

For dilute brushes, $\sigma >>1,$ immersed in a good solvent and
when $\chi $ is independent of$\phi $, $\chi _{eff}(\phi ,T)=\chi
(T)\ll 1/2$, the chemical potential is linear in $\phi $, $\mu
(\phi )\sim \phi $ . In this case, when binary interactions are
dominant, eq. (\ref{profile}) leads to a \textit{parabolic}
concentration profile.\cite{MilnerBrush,Zhulina} At higher
grafting densities, $\sigma \geq 1,$ higher order terms become
significant. These were typically handled by incorporation of the
third virial term.\cite{MilnerBrush,Zhulina} However, as discussed
in the introduction, deviations from these scenarios are expected
when $\chi
_{eff}(\phi )$ varies with $\phi $ and $\mu (\phi )$, as obtained from (\ref%
{Ffint}), assumes the form

\begin{eqnarray}
\mu (\phi )&=&-\ln \left( 1-\phi \right) -1+\chi _{eff}(\phi
)-2\chi _{eff}(\phi )\phi + \nonumber \\
&&\phi (1-\phi )\frac{\partial \chi _{eff}(\phi )}{\partial \phi }
\end{eqnarray}
Since colligative measurements yield $\overline{\chi }(\phi )=\chi
_{eff}(\phi )-(1-\phi )\partial \chi _{eff}(\phi )/\partial \phi $\cite%
{foot2} rather than $\chi _{eff}(\phi )$ it useful to express $\mu (\phi )$
as

\begin{equation}
\mu (\phi )=-\ln (1-\phi )-1+\chi _{eff}(0)-\int_{0}^{\phi }\overline{\chi }%
(\phi )d\phi -\phi \overline{\chi }(\phi )  \label{muint}
\end{equation}

Finally, the Lagrange multiplier $\lambda $ is determined by the
concentration at the outer edge of the brush, $\phi _{H}\equiv \phi (z=H)$

\begin{eqnarray}
\lambda &=&BH^{2}+\mu (z=H)  \nonumber \\
&=&BH^{2}-\ln (1-\phi _{H})-1+\chi _{eff}(0)-\nonumber \\
&&\int_{0}^{\phi _{H}}\overline{%
\chi }(\phi )d\phi -\phi _{H}\overline{\chi }(\phi _{H}),
\end{eqnarray}
In turn, $\phi _{H}$ of a free brush is set by the osmotic pressure at $H$
that is, $\pi _{osm}(\phi _{H})=\left. \phi ^{2}\partial \left[ f_{\infty
}(\phi )/\phi \right] /\partial \phi \right\vert _{\phi =\phi _{H}}=0$,
leading to

\begin{equation}
-\ln (1-\phi _{H})-\phi _{H}-\overline{\chi }(\phi _{H})\phi _{H}^{2}=0
\label{Pi}
\end{equation}
In a good solvent $\phi _{H}=0$.

\section{The Concentration Profiles and Their Moments\label%
{concentrationProf}}

In order to obtain $\phi (z)$ it helpful to express (\ref{profile}) as

\begin{equation}
\Delta \mu (\phi )=B(H^{2}-z^{2})  \label{mustar}
\end{equation}
where

\begin{equation}
\Delta \mu (\phi )=-\ln \frac{1-\phi }{1-\phi _{H}}-\int_{\phi _{H}}^{\phi }%
\overline{\chi }(\phi )d\phi +\phi _{H}\overline{\chi }(\phi _{H})-\phi
\overline{\chi }(\phi )  \label{mu1}
\end{equation}
Equation (\ref{mustar}) does not specify $\phi (z)$ directly. Rather, it
yields $z(\phi )=\sqrt{H^{2}-\Delta \mu (\phi )/B}$. The brush height is
determined in terms of the monomer volume fraction at the surface, $\phi
_{0}=\phi (z=0)$, leading to $H=\sqrt{\Delta \mu (\phi _{0})/B}$ and

\begin{equation}
z(\phi )=\sqrt{\frac{\Delta \mu (\phi _{0})-\Delta \mu (\phi )}{B}}
\label{zphi}
\end{equation}
$\phi (z)$ is determined by equation (\ref{zphi}) together with the
normalization condition (\ref{ConfN1}), which relates $\phi _{0}$ to the
grafting density, $1/\sigma $.

\begin{figure}
\begin{center}
\includegraphics[width=6cm]{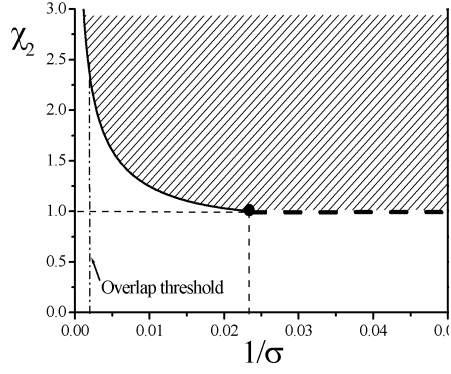}
\end{center}
\caption{The state diagram of a brush with $\overline{%
\chi }(\phi )=1/2+\chi _{2}\phi ^{2}$ and $N=200$ in the $\chi _{2}$, $%
1/\sigma $ plane. Vertical phase separation occurs in the hatched
region. At $\bullet$ $\phi _{0}=\phi_{-}(T)$. The dashed line, at
higher $1/\sigma $, corresponds to brushes exhibiting an
inflection point at altitudes that increase with $1/\sigma $. The
boundary for lower $1/\sigma $ corresponds to $\phi _{-}$ $\phi
_{+}$ coexistence at the grafting surface.} \label{fig2}
\end{figure}

We now distinguish between two cases. In one the concentration profile is
continuous while in the second a discontinuity occurs due to vertical phase
separation. In the first case (\ref{ConfN1}) may be expressed as
\begin{equation}
N=\frac{\sigma }{a^{3}}\int_{\phi _{0}}^{\phi _{H}}\phi \frac{\partial z}{%
\partial \phi }d\phi  \label{sig}
\end{equation}
The concentration profile for a given $\sigma $ is fully specified by (\ref%
{zphi}) and (\ref{sig}). A vertical phase separation in the brush results in
a discontinuity at height $H_{t}$. At this altitude two phases coexist: a
dense inner phase with a monomer volume fraction $\phi _{+}(H_{t})$ and a
dilute outer phase with $\phi _{-}(H_{t})$. In this case the normalization
condition (\ref{ConfN1}) assumes the form
\begin{equation}
N=\frac{\sigma }{a^{3}}\int_{\phi _{0}}^{\phi _{+}(H_{t})}\phi \frac{%
\partial z}{\partial \phi }d\phi +\frac{\sigma }{a^{3}}\int_{\phi
_{-}(H_{t})}^{\phi _{H}}\phi \frac{\partial z}{\partial \phi }d\phi
\label{sigdis}
\end{equation}
where $\phi _{+}(H_{t})$ and $\phi _{-}(H_{t})$ are determined by $\mu (\phi
_{+})=\mu (\phi _{-})$ and $\pi (\phi _{+})=\pi (\phi _{-})$. $\phi (z)$ is
now determined by equation (\ref{zphi}) together with the normalization
condition (\ref{sigdis}),

It is of interest to consider the phase behavior when $\overline{\chi }(\phi
)$ is described by $\overline{\chi }(\phi )=\chi _{0}+\chi _{1}\phi +\chi
_{2}\phi ^{2}$. In the case of $\overline{\chi }(\phi )=\chi _{0}$ or $%
\overline{\chi }(\phi )=\chi _{0}+\chi _{1}\phi $ the critical point, as
specified by $\partial ^{2}f_{\infty }(\phi )/\partial \phi ^{2}=\partial
^{3}f_{\infty }(\phi )/\partial \phi ^{3}=0$

\begin{figure}
\begin{center}
\includegraphics[width=6cm]{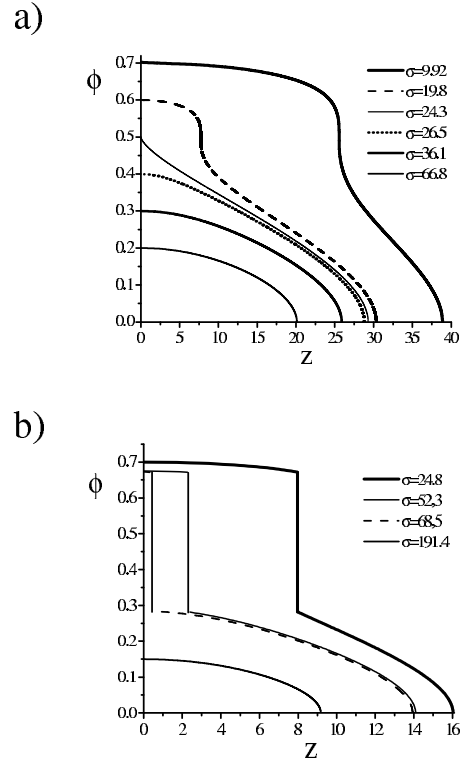}
\end{center}
\caption{$\phi (z)$ for different areas per chain $\sigma $ when
$\overline{\chi }(\phi )=1/2+\chi _{2}\phi ^{2}$. (a) $\chi
_{2}=1$ (b) $\chi _{2}=1.05$. In every case
$N=200$.\cite{FootFig}} \label{fig3}
\end{figure}

\begin{eqnarray}
\frac{1}{1-\phi }-2\overline{\chi }(\phi )-\phi \frac{\partial \overline{%
\chi }(\phi )}{\partial \phi } &=&0  \label{CritP} \\
\frac{1}{\left( 1-\phi \right) ^{2}}-3\frac{\partial \overline{\chi }(\phi )%
}{\partial \phi }-\phi \frac{\partial ^{2}\overline{\chi }(\phi )}{\partial
\phi ^{2}} &=&0  \label{CritP2}
\end{eqnarray}
occurs at $\phi _{c}=0$. This corresponds to the familiar case of a polymer
rich phase in coexistence with a neat solvent. In this situation there is no
vertical phase separation within the brush and the concentration profile is
continuous. A second type of phase separation,\cite{n-Cluster1, n-Cluster2,
Solc1, Solc2} associated with a discontinuous $\phi (z)$, is possible when
higher order terms are involved. For $\overline{\chi }(\phi )=1/2+\chi
_{2}\phi ^{2}$ a critical point occurs at $\phi _{c}=1/2$ and $\chi _{2c}=1$%
. In the vicinity of the critical point, for $\chi _{2}\gtrsim 1$ and $\phi
\gtrsim $ $1/2$ the coexistence curve is well approximated by the spinodal
line $\partial ^{2}f_{\infty }(\phi )/\partial \phi ^{2}=0$

\begin{equation}
\frac{1}{1-\phi }-1-4\chi _{2}\phi ^{2}=0  \label{spinodal}
\end{equation}
leading to $\phi _{\pm }=\frac{1}{2}\pm \frac{1}{2}\sqrt{1-1/\chi _{2}}$.
The state diagram of a brush in the $\chi _{2},1/\sigma $ plane when eq. (%
\ref{spinodal}) applies is shown in Fig. \ref{fig2}. Concentration profiles
obtained from $\overline{\chi }(\phi )$ of this form are depicted in Fig. %
\ref{fig3}.

\begin{figure}
\begin{center}
\includegraphics[width=6cm]{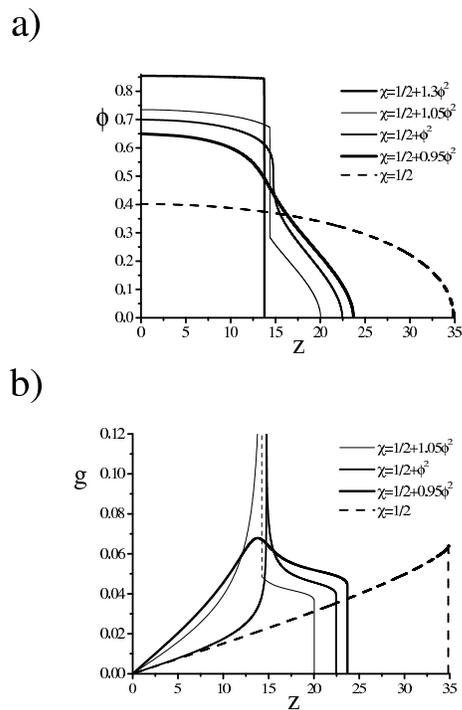}
\end{center}
\caption{Plots of $\phi (z)$ (a) and $g(z)$ (b) above and
below the critical point for $\overline{\chi }(\phi )=1/2+\chi _{2}\phi ^{2}$%
, $\sigma =17$ and $N=200 $. In all cases, the outer phase is swollen.\cite%
{FootFig}} \label{fig4}
\end{figure}

Experimentally, the brush thickness $H$ is inaccessible. Certain
experimental technique yield $\phi (z)$ allowing one to obtain moments of $%
\phi (z)$

\begin{equation}
\langle z\rangle =\frac{\int_{0}^{H}z\phi (z)dz}{\int_{0}^{H}\phi (z)dz}=%
\frac{\sigma }{Na^{3}}\int_{0}^{H}z\phi (z)dz  \label{firstmom}
\end{equation}

\begin{equation}
\langle z^{2}\rangle =\frac{\int_{0}^{H}z^{2}\phi (z)dz}{\int_{0}^{H}\phi
(z)dz}=\frac{\sigma }{Na^{3}}\int_{0}^{H}z^{2}\phi (z)dz.  \label{secondmom}
\end{equation}%
Other techniques, such as ellipsometry, measure $\langle z\rangle
$.\cite{charmet} As we shall discuss, the $\sigma $ dependence of
the moments provides useful information on the brush structure.
The details of the calculation of these moments are described in
Appendix \ref{appendixCalc}.

When $\phi (z)$ is continuous, both moments increase smoothly with the
grafting density. In marked contrast, vertical phase separation within the
brush gives rise to a non-monotonic behavior. In particular, both $\langle
z\rangle $ and $\sqrt{\langle z^{2}\rangle }$ exhibit a minimum at
intermediate $\sigma $. A vertical phase separation gives rise to a plateau
in the $H$ vs. $\sigma $ plot (Fig. \ref{fig8}) while in the plots of $%
\langle z\rangle $ and $\sqrt{\langle z^{2}\rangle }$ vs. $\sigma $ it is
associated with a minimum (Fig. \ref{fig6} and Fig. \ref{fig7}). The
physical origin of this behavior is the partitioning of the monomers between
the inner dense phase and the outer dilute one. The minima are traceable to
the higher weight give to the inner phase. Since the inner phase is denser,
the onset of vertical phase separation is associated with a decrease $%
\langle z\rangle $ and $\sqrt{\langle z^{2}\rangle }$. These
features provide a useful diagnostic for the occurrence of a
vertical phase separation in the brush. The SCF analysis in this
section confirms earlier results\cite{BH2} obtained by utilizing
the Pincus approximation.\cite{Pincus, Pincus1} As we shall
discuss this is the case for properties that are insensitive to
the precise form of $g(z)$. In marked contrast, the compression
force profile (section V) does depend on $g(z)$ and the SCF result
differ from the one obtained from the Pincus approximation.

\begin{figure}
\begin{center}
\includegraphics[width=6cm]{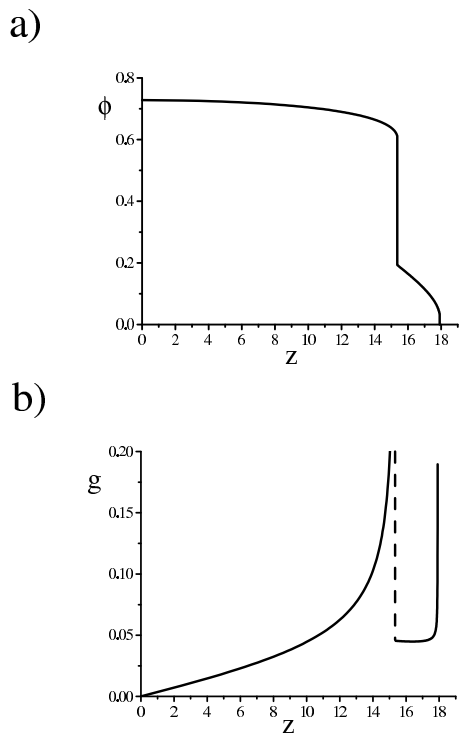}
\end{center}
\caption{Plots of $\phi (z)$ (a) and $g(z)$ (b) for the case of
two coexisting dense phases with $\overline{\chi }(\phi
)=0.51+\chi _{2}\phi ^{2}$, $\sigma =18$ and
$N=200$.\cite{FootFig}} \label{fig5}
\end{figure}

\section{The Distribution of Free Ends\label{distribution}}

The SCF formalism allows to obtain the height distribution of free ends, $%
g(z)$. Current experimental techniques do not allow to probe $g(z)$
directly. However, $g(z)$ is of interest because it plays a role in the
calculation of the compression force profile. When $\chi =const$ the brush
structure is dominated by the contributions of the second and third virial
terms of $F_{int}$. Three scenarios emerge. In a good solvent the ends are
distributed throughout the brush and $g(z)$ is a smooth function vanishing
at $z=0$ and $z=H$. When the brush is collapsed in a poor solvent the ends
reside preferentially at the outer edge of the brush and $g(z)$ diverges at $%
H$. In a $\theta $ solvent $g(z)$ increases smoothly with $z$ but does not
diverge.\cite{Skvortsov,Zhulina} As we shall see, a new scenario emerges
when a vertical phase separation occurs. In particular, $g(z)$ will then
diverge at the phase boundary indicating localization of the ends at the
boundary. We will obtain $g(z)$ from the integral equation (\ref{ConFi}).
The details of the calculation are described in Appendix \ref{appendixg}.

\begin{figure}
\begin{center}
\includegraphics[width=8cm]{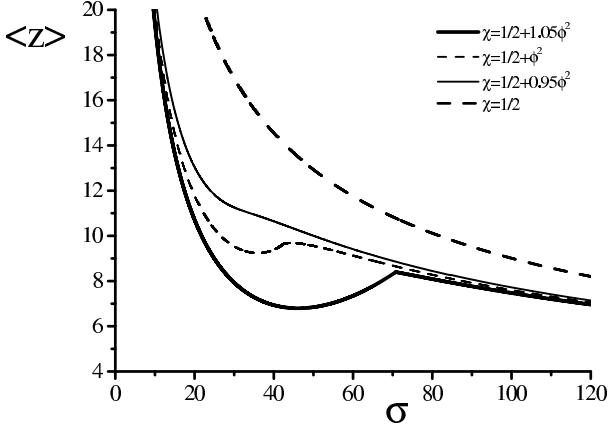}
\end{center}
\caption{$\langle z\rangle $ as a function of $\sigma $ for
different $\chi _{2}$ values and $N=300$.\cite{FootFig}}
\label{fig6}
\end{figure}

$g(z)$ of a brush with a continuous profile (Fig. \ref{fig4}a) is specified
by

\begin{eqnarray}
g(z)&=&z\frac{\sigma }{Na^{3}}\left( \frac{\phi
_{H}}{\sqrt{H^{2}-z^{2}}}+\right. \nonumber \\
&&\left. \sqrt{B}\int_{\phi _{H}}^{\phi }\frac{d\phi ^{\prime
}}{\sqrt{\Delta \mu (\phi )-\Delta \mu (\phi ^{\prime })}}\right)
\label{gCon}
\end{eqnarray}
where $\phi $ and $z$ are related by (\ref{mustar}). When a vertical
separation occurs within the brush, equation (\ref{ConFi}) yields now two
expressions, for $\phi (z)$. At the outer edge, $H_{t}< z< H$

\begin{equation}
\phi (z)=\frac{a^{3}}{\sigma }\int_{z}^{H}\frac{g(z^{\prime })dz^{\prime }}{%
E(z,z^{\prime })}  \label{phim}
\end{equation}
while at the inner dense phase, $0< z< H_{t}$

\begin{equation}
\phi (z)=\frac{a^{3}}{\sigma }\int_{z}^{H_{t}}\frac{g(z^{\prime })dz^{\prime
}}{E(z,z^{\prime })}+\frac{a^{3}}{\sigma }\int_{H_{t}}^{H}\frac{g(z^{\prime
})dz^{\prime }}{E(z,z^{\prime })}  \label{phip}
\end{equation}
In the outer region only free ends with $H_{t}< z$ contribute while for the
inner phase all free ends are involved.

The expression for $g(z)$ in the two regions are given below while the
details of the derivation are presented in appendix \ref{appendixg}. At the
outer phase, $H_{t}< z< H$

\begin{eqnarray}
g(z)&=&z\frac{\sigma }{Na^{3}}\left( \frac{\phi
_{H}}{\sqrt{H^{2}-z^{2}}}+\right. \nonumber \\
&& \left. \sqrt{B}\int_{\phi _{H}}^{\phi }\frac{d\phi ^{\prime
}}{\sqrt{\Delta \mu (\phi )-\Delta \mu (\phi ^{\prime })}}\right)
\label{gm}
\end{eqnarray}
while in the inner phase, $0< z< H_{t}$
\begin{eqnarray}
&&g(z)=\frac{z\sigma }{Na^{3}}\left[ \frac{\phi _{+}(H_{t})-\phi _{-}(H_{t})%
}{\sqrt{H_{t}^{2}-z^{2}}}+\frac{\phi _{H}}{\sqrt{H^{2}-z^{2}}}+\right. \nonumber \\
&&\sqrt{B}\int_{\phi _{+}(H_{t})}^{\phi }\frac{d\phi ^{\prime
}}{\sqrt{\Delta \mu
(\phi )-\Delta \mu (\phi ^{\prime })}}+\frac{\sqrt{B}}{\pi }\sqrt{H_{t}^{2}-z^{2}}\times  \nonumber \\
&& \int_{\phi _{H}}^{\phi _{\_}(H_{t})}\frac{\frac{d\Delta \mu
(\phi
^{\prime })}{d\phi ^{\prime }}d\phi ^{\prime }}{\left( H^{2}-z^{2}-\frac{%
\Delta \mu (\phi ^{\prime })}{B}\right)
\sqrt{H^{2}-H_{t}^{2}-\frac{\Delta \mu (\phi ^{\prime })}{B}}}
\times \nonumber \\
&&\left. \int_{\phi _{H}}^{\phi ^{\prime }}\frac{d\phi ^{\prime
\prime }}{\sqrt{\Delta \mu (\phi ^{\prime })-\Delta \mu (\phi
^{\prime \prime })}}\right] .  \label{gp(z)}
\end{eqnarray}
The first integral allows for the contribution of the inner phase and the
second for the contribution of the outer phase. $g(z)$ (\ref{gp(z)}) at the
interval $0<z<H_{t}$ diverges at the phase boundary $z=H_{t}$. $g(z)$ (\ref%
{gm}) at the interval $H_{t}<z<H$ diverges at $H$ when the outer phase is
collapsed and $\phi _{H}>0$. In this case the two coexisting phases are
dense (Fig. \ref{fig5}). When $\phi _{H}=0$ the outer phase is swollen and $%
g(z)$ does not diverge at $H$ (Fig. \ref{fig4}b). A rough approximation
yielding closed form expressions for $g(z)$ for discontinuous brushes is
described in Appendix \ref{appendixModel}.

\begin{figure}
\begin{center}
\includegraphics[width=8cm]{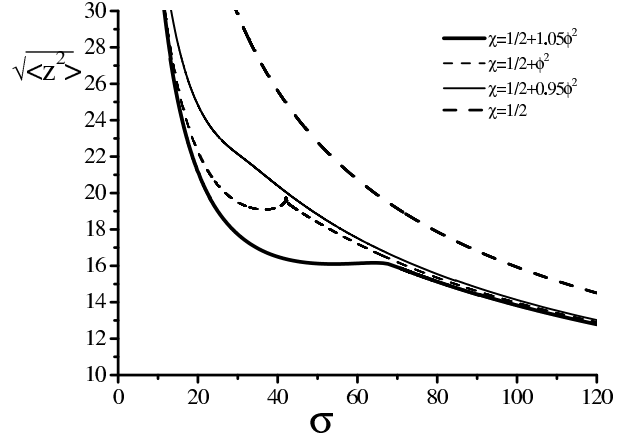}
\end{center}
\caption{$\sqrt{\langle z^{2}\rangle}$ as a function of $%
\sigma $ for different $\chi _{2}$ values and
$N=300$.\cite{FootFig}} \label{fig7}
\end{figure}

\section{The Compression Force Profile of a "Two-State" Brush\label%
{compression}}

The surface force apparatus allows to measure the restoring force arising
upon compression of a brush. For brushes of polymers characterized by a
constant $\chi $ the force increases smoothly with the compression and the
force profile is essentially featureless. When the brush consists of
polymers characterized by $\overline{\chi }(\phi )$ the compression can
induce a vertical phase separation even if the concentration profile of the
brush is initially continuous. The existence of vertical phase separation,
be it compression induced or not, gives rise to distinctive regimes in the
force profile. In particular, the slope of the force vs. distance curve in
different compression regimes can be markedly different. In such experiments
$H$ is determined by the compressing surface rather than by $\sigma $.
Accordingly, $\phi _{H}$ is set by the normalization condition (\ref{ConfN1}%
) and not by $\pi (\phi _{H})=0$. The compression increases
$F_{chain}$ and the restoring force per area $\sigma $ is
\begin{equation}
f(H)=-\frac{\partial F_{chain}}{\partial H}  \label{Force}
\end{equation}%
In the following we obtain this force law for the case of compression by
impenetrable, non-adsorbing surface.

\begin{figure}
\begin{center}
\includegraphics[width=7cm]{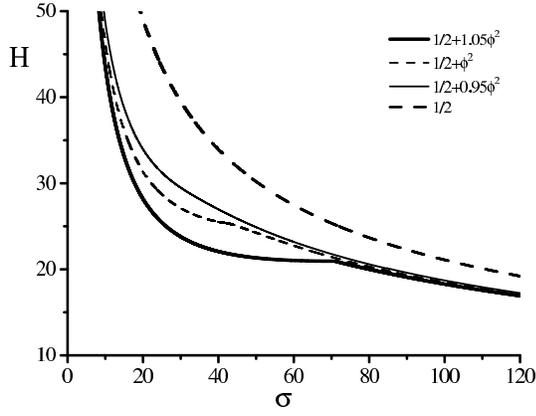}
\end{center}
\caption{$H$ as a function of $\sigma $ for different $%
\chi _{2}$ values and $N=300$.\cite{FootFig}} \label{fig8}
\end{figure}

For a brush with a continuous $\phi (z)$

\begin{equation}
\frac{F_{chain}}{kT}=\int_{\phi _{0}}^{\phi _{H}}\left[ \frac{\sigma }{a^{3}}%
f_{\infty }(\phi )+\frac{3}{2a^{2}}\frac{\pi ^{2}}{8N}z^{2}(\phi )g(\phi )%
\right] \frac{\partial z}{\partial \phi }d\phi
\end{equation}
obtained by invoking $\int_{0}^{z^{\prime }}E(z^{\prime },z)dz=z^{\prime
2}\pi ^{2}/(8N)$. Here $g(z)$ is given by (\ref{gCon}), while $z(\phi )$ and
$\partial z/\partial \phi $ are specified by (\ref{zphi}). $f(H)$ is
calculated numerically subject to the constraint (\ref{sig}). When the
concentration at the wall, $\phi _{0}$, exceeds $\phi _{-}$, the brush
undergoes a vertical phase separation and $\phi (z)$ is no longer
continuous. In this case $F_{chain}$ assumes the form

\begin{figure}
\begin{center}
\includegraphics[width=7cm]{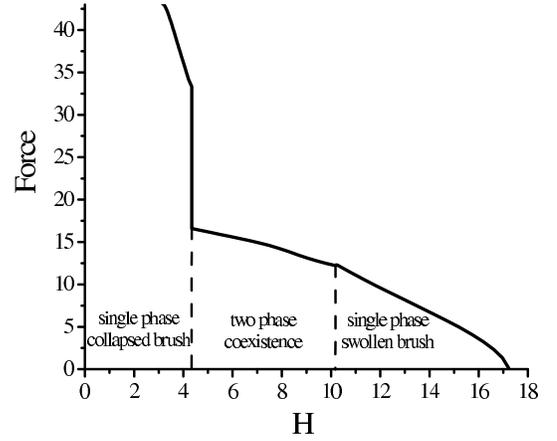}
\end{center}
\caption{The compression force profile for a brush with $%
\overline{\chi }(\phi )=1/2+1.05\phi ^{2}$, $\sigma =120$ and
$N=300$. The
uncompressed brush is in a single phase state ($\phi _{0}<\phi _{-})$.\cite%
{FootFig}} \label{fig9}
\end{figure}

\begin{eqnarray}
\frac{F_{chain}}{kT} &=&\frac{\sigma }{a^{3}}\int_{\phi
_{0}}^{\phi _{H}}f_{\infty }(\phi )\frac{\partial z}{\partial \phi
}d\phi +\nonumber \\
&&\frac{3}{2a^{2}}\frac{\pi ^{2}}{8N}\left[ \int_{\phi _{0}}^{\phi
_{+}(H_{t})}z^{2}(\phi )g(\phi )\frac{\partial z}{\partial \phi }d\phi +\right. \nonumber \\
&& \left. \int_{\phi _{-}(H_{t})}^{\phi _{H}}z^{2}(\phi )g(\phi
)\frac{\partial z}{\partial \phi }d\phi \right]
\end{eqnarray}
where $g(\phi )$ is specified by eqs. (\ref{gp(z)}) for the inner phase and
by (\ref{gm}) for the outer one. The conservation of monomers is enforced by
the constraint (\ref{sigdis}).

\begin{figure}[tbh]
\begin{center}
\includegraphics[width=3.2cm]{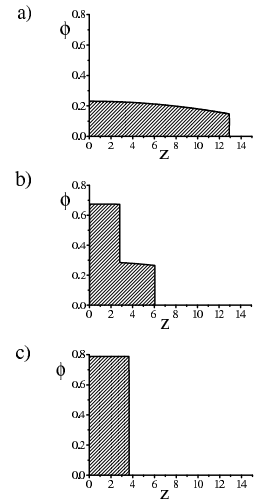}
\end{center}
\caption{$\phi (z)$ plots corresponding to the three regimes in
Fig. \ref{fig9}. (a) single phase swollen phase, $H=12.9$, (b) a
coexistence of a dense and a dilute phase, $H=6.1$, (c) single
dense phase, $H=3.7$. In every case $\sigma =120$ and
$N=300$.\cite{FootFig}} \label{fig10}
\end{figure}

When the conditions permit a vertical phase separation within the brush, it
can take place in two ways. It can occur when the grafting density exceeds a
certain critical value thus causing $\phi _{0}>\phi _{-}$. Alternatively, it
can also take place as a result of compression when the grafting density
does not lead to phase separation in the unperturbed brush. The development
of $\phi (z)$ and $f(H)$ for this second case is depicted in figures \ref%
{fig9} and \ref{fig10} respectively. Initially, the brush retains the single
phase structure and the associated force law. When the compression enforces $%
\phi _{0}>\phi_{-}$, a vertical phase separation occurs and is signalled by
a weaker slope of the $f(H)$ vs. $H$ curve. Stronger compression causes
complete conversion to a dense phase thus causing an abrupt increase in $%
f(H) $.

\section{ The Karlstrom Model and the distribution of monomeric states \label%
{distributionKarlsrom}}

Thus far, our discussion concerned brushes characterized by an arbitrary $%
\chi _{eff}(\phi )$. We now illustrate these considerations for the case of
the Karlstrom model.\cite{Karlstrom} We focus on this model because of its
simplicity and its semiquantiative agreement with the phase diagram of
aqueous solutions of PEO at atmospheric pressure\cite{Karlstrom} and the
measured $\overline{\chi }(T,\phi )$.\cite{BH} Within this model the
monomers exist in two states: a polar, hydrophilic state (A) and an apolar,
hydrophobic state (B). The two intercovert via internal rotations. In the $%
N\rightarrow \infty $ limit the corresponding Flory-type free energy is
\begin{eqnarray}
\frac{f_{\infty }(\phi )}{kT} &=&(1-\phi )\ln (1-\phi )+ \nonumber \\
&& \phi \lbrack p\ln p+(1-p)\ln
(1-p)]+  \nonumber \\
&&\phi p\Delta \epsilon +\phi (1-\phi )[p\chi _{AS}+(1-p)\chi
_{BS}]+\nonumber \\
&&\phi ^{2}\chi _{AB}p(1-p)  \label{K}
\end{eqnarray}
where $p$ denotes the fraction of monomers in the A state. The first term
allows for the mixing entropy of the solvent while the second reflects the
mixing entropy of the A and B states along the chain. $\Delta \epsilon $ is
the energy difference between two states and $\phi p\Delta \epsilon $ allows
for the effect of interconversion between them. The fourth term is the
generalization of the $\chi \phi (1-\phi )$ in the Flory free energy to
allow for the interactions of the two states with the solvent S. The
interactions between $A$ and $B$ states gives rise to the last term.

The equilibrium value of $p$ for a given $\phi $ is specified by $\partial
f_{\infty }/\partial p=0$ leading to
\begin{eqnarray}
\frac{p}{1-p}&=&\exp \left[ -\Delta \epsilon -(1-\phi )(\chi
_{AS}-\chi _{BS})-\right. \nonumber \\
&& \left. \phi \chi _{AB}(1-2p)\right] \label{eqP}
\end{eqnarray}
The parameters used by Karlstrom\cite{Karlstrom} to fit phase diagram of PEO
in water are $\chi _{As}=80.0/T$, $\chi _{Bs}=684.5/T$, $\chi _{AB}=155.6/T$%
, $\Delta \varepsilon =-625.2/T+\ln 8$. In the remainder of this section we
consider brushes at $T=60^{o}$ C. The equilibrium $\chi _{eff}(\phi )$ is
obtained by equating (\ref{K}) and (\ref{eqP})

\begin{eqnarray}
\chi _{eff}(\phi ) &=&p\chi _{AS}+(1-p)\chi _{BS}+  \nonumber \\
&&\frac{\phi }{1-\phi }\left[ \chi _{AB}p(1-p)+
p\Delta \epsilon \right] + \nonumber \\
&& \frac{p\ln p+(1-p)\ln (1-p)}{1-\phi }  \label{chieff}
\end{eqnarray}
where $p(\phi )$ is specified by (\ref{eqP}). $\ \overline{\chi }=-\frac{%
\partial }{\partial \phi }\frac{f_{\infty }(\phi )}{\phi }$ together with (%
\ref{eqP}) yield the equilibrium value
\begin{equation}
\overline{\chi }(\phi )=p\chi _{AS}+(1-p)\chi _{BS}-\chi _{AB}p(1-p)
\label{chiK}
\end{equation}
Both $\chi _{eff}(\phi )$ and $\overline{\chi }(\phi )$ can be expanded in
powers of $\phi $. The coefficients in the expansion depend on the
parameters, $\Delta \epsilon $, $\chi _{AS}$, $\chi _{BS}$, $\chi _{AB}$.
High order terms are of negligible importance and $\overline{\chi }(\phi
)\approx 0.48+0.31\phi +0.07\phi ^{2}$ provides a good approximation for $%
\overline{\chi }(\phi )$.

\begin{figure}
\begin{center}
\includegraphics[width=8cm]{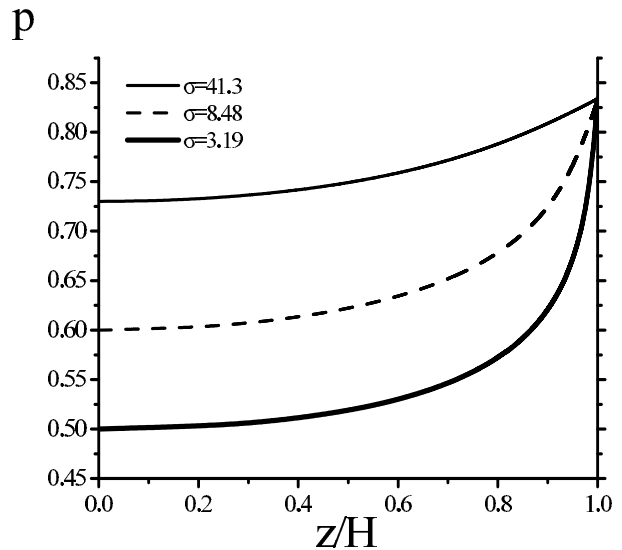}
\end{center}
\caption{A plot of the fraction of hydrophilic states, $%
p $, vs. $z$, in the Karlstrom model\cite{Karlstrom} for different $\sigma $%
. In every case $\chi _{As}=80.0/T$, $\chi _{Bs}=684.5/T$, $\chi
_{AB}=155.6/T$, $\Delta \varepsilon =-625.2/T+\ln 8$, $N=300$
$T=60^{o}$ C. \cite{FootFig}} \label{fig11}
\end{figure}

\begin{figure}
\begin{center}
\includegraphics[width=8cm]{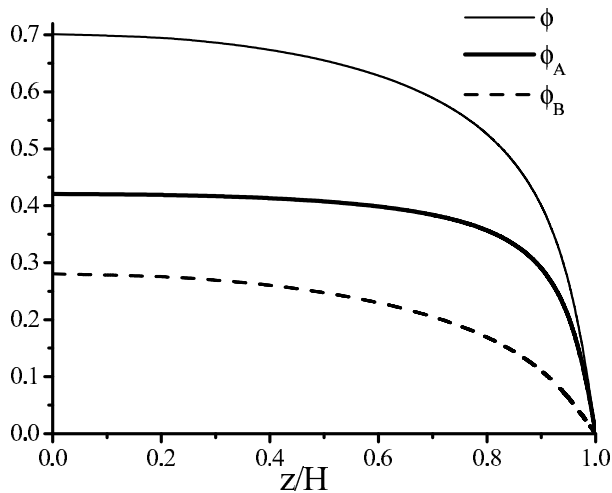}
\end{center}
\caption{The overall $\phi (z)$ and the corresponding
concentration profiles of the monomeric states $\phi
_{A}(z)=p(z)\phi (z)$ and $\phi _{B}(z)=[1-p(z)]\phi (z)$ within
the Karlstrom model for the conditions specified in Fig.
\ref{fig11}.\cite{FootFig}} \label{fig12}
\end{figure}

Equation (\ref{eqP}) allows to relate the volume fraction $\phi $ to $p$,
the fraction of hydrophilic A states, as

\begin{equation}
\phi (p)=\frac{\ln \frac{p}{1-p}+\Delta \epsilon +\chi _{AS}-\chi _{BS}}{%
\chi _{AS}-\chi _{BS}-\chi _{AB}(1-2p)}
\end{equation}
Accordingly, the exchange chemical potential can be specified in terms of $p$%
, \textit{i.e. }$\Delta \mu (\phi (p))$. In turn, eq. (\ref{mustar}) enables
us to obtain $z=z(p)$. To this end we invoke two boundary conditions: (i) In
this range of parameters the brush is swollen and $\phi $ vanishes at the
outer edge, $\phi _{H}=\phi (p=p_{H})=0$ where $p_{H}$ is the value of $p$
at the height $H$. (ii) At the grafting surface we have $\Delta \mu
(z=0,p=p_{0})=BH^{2}$, where $p_{0}$ is the value of $p$ at $z=0$. In
addition we utilize the conservation of monomers as given by (\ref{sig}).
The corresponding plots of $p=p(z)$ as well as the concentration profiles of
the two states are depicted, for different $\sigma $, in Fig. \ref{fig11}
and in Fig. \ref{fig12}. Since all brushes considered are swollen, with $%
\phi _{H}=0$, the $p$ values at the outer edge of the brush, $z=H$ are
identical, $p=p_{H}$. Increasing grafting density leads to higher
concentration at the grafting surface. This favors the hydrophobic B state
and lower $p$ at the surface. For the chosen parameters, the minimal value
of $p$, corresponding to a PEO melt ($\phi =1$), is $p_{\ast }=0.45$.

\section{Discussion\label{discussion}}

In this article we presented a common framework for the analysis of the
structure of brushes of neutral water-soluble polymers (NWSP) that own their
solubility to the formation of H-bonds with water molecules. Our analysis
concerned a family of two-state models developed for PEO but applicable, in
principle, to other NWSP. The particular aspects of the models were grouped
into $\chi _{eff}(T,\phi )$ thus allowing for a unified discussion of the
brush structure within these models. Significant part of the discussion
concerned brushes exhibiting vertical phase separation that can occur for
polymers capable of a second type of phase separation. In particular, we
examined the distinctive behavior of plots of $\langle z\rangle $ and $\sqrt{%
\langle z^{2}\rangle }$ vs. $\sigma $ and the compression force
profiles associated with such brushes. $\phi (z)$ and its moments
are insensitive to the precise form of $g(z)$ and the SCF analysis
recovers the results obtained earlier\cite{BH2} by the use of the
Pincus approximation.\cite{Pincus,Pincus1} In marked contrast, the
compression force law does depend on $g(z)$ and a full SCF
analysis is necessary in order to obtain the correct results.
These features are useful criteria for the occurrence of vertical
phase separation. Such criteria are of interest because of
indirect experimental indications that brushes of PNIPAM exhibit
this effect. Among these indications the following are especially
noteworthy. First, is an early study by Zhu and
Napper\cite{Napper} of the collapse of PNIPAM brushes grafted to
latex particles immersed in water. This revealed a collapse
involving two stages. An \textquotedblleft early
collapse\textquotedblright
, took place below $30^{o}$ C, at \textquotedblleft better than $\theta $%
-conditions\textquotedblright , and did not result in flocculation of the
neutral particles. Upon raising the temperature to \textquotedblleft worse
than $\theta $-conditions\textquotedblright\ the collapse induced
flocculation. This indicates that the colloidal stabilization imparted by
the PNIPAM brushes survives the early collapse. It lead to the
interpretation of the effect in terms of a vertical phase separation within
the brush associated with a second type of phase separation as predicted by
the $n$-cluster model. More recently, Balamurugan \textit{et al}\cite%
{Balamurugan} used Surface Plasmon Resonance (SPR) and contact
angle measurements with water to characterize PNIPAM brushes
\textquotedblright grafted from\textquotedblright\ a
self-assembled monolayer on gold. The brush properties were
studied at between $10^{o}$ C and $40^{o}$ C. The SPR measurements
indicated gradual variation of the brush properties. In marked
contrast, the contact angle measurements revealed an abrupt change
at $\sim 32^{o}$ C. Both experiments are consistent with a brush
undergoing vertical phase separation such that the outer phase is
hydrophilic. Within this picture, the abrupt changes reported in
the wetting and aggregation behavior correspond than to the onset
of a single, dense and hydrophobic phase. This picture is also
supported by recent study of the phase behavior of PNIPAM by
Afroze \textit{et al}.\cite{A} Early study of the phase behavior
of PNIPAM in water, by Heskins and Guillet,\cite{HG} identified a
LCST at $\phi _{c}\simeq 0.16$ and $T_{c}\simeq 31.0^{o}$ C. In
marked contrast, the work of Afroze \textit{et al}\cite{A}
identified PNIPAM as a polymer undergoing a second type of phase
separation. In particular: (i) While the LCST of PNIPAM depends on
$N$ the LCST occurs around $T_{c}\simeq 27-28^{o}$ C and $\phi
_{c}\simeq 0.43$ (ii) In the limit of $\phi \rightarrow 0$, the
phase separation occurs, depending on $N$, between $30^{o}$ C and
$34^{o}$ C. Thus, the phase diagram of Afroze \textit{et al
}suggests that a vertical phase separation is indeed expected in
brushes of PNIPAM.\cite{BH2} Systematic Neutron Reflectometry (NR)
studies of PNIPAM brushes will eventually provide clearer picture
of the situation. Early studies were hampered by high
polydispersity as well as difficulties in determining $N$ and
$\sigma $.\cite{Yim} With this in mind, the NR results revealed
that the structure of PNIPAM brushes in acetone was very different
from their structure in water, both at $20^{o}$ C and at $55^{o}$
C. In acetone the concentration profile was smoothly decaying
while in water it consisted of a narrow, inner, dense region and
an outer, extended and dilute region. More recent work utilized
NR,\cite{Yim1} to study samples with lower polydispersity and
higher grafting density between $20^{o}$ C and $40^{o}$ C.
Importantly, the results indicate that most of the conformational
change occurred between $28^{o}$ C and $34^{o}$ C though the
corresponding concentration profiles were not reported. The
results suggest a repartitioning of the monomers between a dilute
outer tail and an inner dense region.\cite{Yim2} Finally, recent
experimental results of Hu \textit{at al}\cite{Hurec} are
suggestive of the predictions obtained above concerning the
variation of $\langle z\rangle$ upon decreasing $\sigma$ (Fig.
\ref{fig6}). Hu \textit{et al} studied the thickness of a PNIPAM
brush grafted to spherical microgels of copolymers of PNIPAM and
acrylic acid 2-hydroxyethyl ester (HEA). The microgels shrink as
$T$ increases from $24^o$ C to $36^o$ C thus inducing a decrease
in $\sigma$. Remarkably, the thickness of the brush initially
decreases in the range $27^o$ C to $32^o$ C but subsequently
increases upon further heating in the range $27^o$--$32^o$ C.
Unfortunately, in this experiment it is impossible to separate the
effects due to change in $T$ from those to the change in $\sigma$
because the $T$ is used to tune $\sigma$.

In confronting experimental results with theoretical predictions concerning
the vertical phase separation it is important to note two points. First,
polydispersity in $N$ and in $\sigma $ can smooth out the discontinuity in $%
\phi (z)$. These factors give rise to domains with different $\phi _{0}$. In
turn, the altitude of the discontinuity in these domains will also differ.
NR measurements will reflect the weighted average of these domains.
Polydispersity in $N$ can lead to further strengthening of this effect if $%
\chi _{eff}$ depends on $N$. Second, the vertical phase separation involves
a first order phase transition. Accordingly, nucleation dynamics may play a
role and it is necessary to allow for the possibility of long relaxation
times.

\appendix

\section{Calculation of the average thickness\label{appendixCalc}}

For a $\phi (z)$ (\ref{firstmom}) is evaluated using integration by parts
utilizing $d\Delta \mu =-2Bzdz$:

\begin{eqnarray}
\langle z\rangle &=&\frac{\sigma
}{Na^{3}}\frac{1}{2B}\int_{0}^{BH^{2}}\phi (z)d\Delta \mu
\nonumber \\
&=&\frac{\sigma }{Na^{3}}\left( H^{2}\frac{\phi _{0}}{2}-\frac{1%
}{2B}\int_{\phi _{H}}^{\phi _{0}}\Delta \mu (\phi )d\phi \right)
\label{mom1}
\end{eqnarray}
For a discontinuous $\phi (z)$ (\ref{firstmom}) this procedure leads to

\begin{eqnarray}
\langle z\rangle &=&\frac{\sigma }{Na^{3}}\left[ \int_{0}^{H_{t}}z\phi
(z)dz+\int_{H_{t}}^{H}z\phi (z)dz\right]  \nonumber \\
&=&\frac{\sigma }{Na^{3}}\frac{1}{2B}\left(
\int_{B(H^{2}-H_{t}^{2})}^{BH^{2}}\phi (z)d\Delta \mu
+ \right. \nonumber \\
&&\left. \int_{0}^{B(H^{2}-H_{t}^{2})}\phi (z)d\Delta \mu \right)  \nonumber \\
&=&\frac{\sigma }{Na^{3}}[H^{2}\frac{\phi _{0}}{2}-\left(
H^{2}-H_{t}^{2}\right) \frac{\phi _{+}(H_{t})-\phi _{-}(H_{t})}{2}  \nonumber
\\
&-&\frac{1}{2B}\int_{\phi _{+}(H_{t})}^{\phi _{0}}\Delta \mu (\phi )d\phi -%
\frac{1}{2B}\int_{\phi _{H}}^{\phi _{-}(H_{t})}\Delta \mu (\phi
)d\phi ] \label{mom11}
\end{eqnarray}

The evaluation of (\ref{secondmom}) in the case of a continuous $\phi (z)$
involves introducing the variable $\Delta \mu =B(H^{2}-z^{2})$ and invoking $%
z=\sqrt{H^{2}-\Delta \mu /B}$. Integration by parts leads to

\begin{eqnarray}
\langle z^{2}\rangle &=&\frac{\sigma }{Na^{3}}\frac{1}{2B}
\int_{0}^{BH^{2}}z\phi (z)d\Delta \mu \nonumber \\
&=&-\frac{\sigma }{Na^{3}}\frac{1}{2B}\int_{\Delta \mu =0}^{\Delta
\mu =BH^{2}}\Delta \mu d(z\phi (z)) \nonumber \\
&=&-\frac{\sigma }{Na^{3}}\frac{1}{2B}\int_{\phi _{H}}^{\phi
_{0}}\left( \phi \frac{\partial z}{\partial \phi }+z\right) \Delta
\mu d\phi \label{mom2}
\end{eqnarray}
Using

\begin{equation}
\frac{\partial z}{\partial \phi }=-\frac{1}{B\sqrt{H^{2}-\Delta \mu /B}}%
\frac{\partial \Delta \mu }{\partial \phi }
\end{equation}
we obtain the final expression

\begin{eqnarray}
\langle z^{2}\rangle &=&\frac{\sigma
}{Na^{3}}\frac{1}{2B^{2}}\int_{0}^{\phi
_{0}}\frac{\phi \Delta \mu \frac{\partial \Delta \mu }{\partial \phi }}{%
\sqrt{H^{2}-\Delta \mu /B}}d\phi -\nonumber \\
&&\frac{1}{2B}\int_{0}^{\phi _{0}}\Delta \mu \sqrt{H^{2}-\Delta
\mu /B}d\phi  \label{mom22}
\end{eqnarray}
where $\Delta \mu $ is specified by (\ref{mu1}).

Following a similar procedure for a discontinuous $\phi (z)$ (\ref{secondmom}%
) yields

\begin{eqnarray}
\langle z^{2}\rangle &=&\frac{\sigma }{Na^{3}}\left[ \int_{0}^{H_{t}}z^{2}%
\phi (z)dz+
\int_{H_{t}}^{H}z^{2}\phi (z)dz\right]  \nonumber \\
&=&\frac{\sigma }{Na^{3}}\frac{1}{2B}\left(
\int_{B(H^{2}-H_{t}^{2})}^{BH^{2}}z\phi (z)d\Delta \mu
+ \right. \nonumber \\
&& \left. \int_{0}^{B(H^{2}-H_{t}^{2})}z\phi (z)d\Delta \mu \right)  \nonumber \\
&=&\frac{\sigma }{Na^{3}}\left[ \frac{1}{2B^{2}}\int_{\phi
_{+}(H_{t})}^{\phi _{0}}\frac{\phi \Delta \mu \frac{\partial \Delta \mu }{%
\partial \phi }}{\sqrt{H^{2}-\Delta \mu /B}}d\phi - \right.\nonumber \\
&&\frac{1}{2B}\int_{\phi _{+}(H_{t})}^{\phi _{0}}\Delta \mu
\sqrt{H^{2}-\Delta \mu /B}d\phi
\nonumber \\
&&+\frac{1}{2B^{2}}\int_{0}^{\phi _{-}(H_{t})}\frac{\phi \Delta \mu \frac{%
\partial \Delta \mu }{\partial \phi }}{\sqrt{H^{2}-\Delta \mu /B}}d\phi
- \nonumber \\
&&\frac{1}{2B}\int_{0}^{\phi _{-}(H_{t})}\Delta \mu \sqrt{H^{2}-\Delta \mu /B}%
d\phi  \nonumber \\
&&\left. -H_{t}\left( H^{2}-H_{t}^{2}\right) \frac{\phi _{+}(H_{t})-\phi
_{-}(H_{t})}{2}\right]  \label{mom222}
\end{eqnarray}

\section{Calculation of the distribution of ends function\label{appendixg}}

Upon introducing the variables $\rho =H^{2}-z^{2}$, $t=H^{2}-z^{\prime 2}$
and $g(z^{\prime })dz^{\prime }=-f(t)dt$ eq. (\ref{ConFi}) assumes the form
of an Abel integral equation

\begin{equation}
v(\rho )=\int_{0}^{\rho }\frac{f(t)dt}{\sqrt{\rho -t}},  \label{Abel}
\end{equation}
where $v(\rho )=\frac{\pi \sigma }{2Na^{3}}\phi (z)$ whose solution (\ref%
{Abel}) is

\begin{equation}
f(\rho )=\frac{1}{\pi }\left( \frac{v(0)}{\sqrt{\rho }}+\int_{0}^{\rho }%
\frac{1}{\sqrt{\rho -t}}\frac{dv(t)}{dt}dt\right) .  \label{solAbel}
\end{equation}
It is convenient to rewrite the integral in (\ref{solAbel}) as

\begin{eqnarray}
&&\int_{0}^{\rho }\frac{1}{\sqrt{\rho -t}}\frac{dv(t)}{dt}dt=\int_{v(H)}^{v(z)}%
\frac{dv}{\sqrt{\rho -t(v)}}\nonumber \\
&=&\frac{\pi \sigma
}{2Na^{3}}\sqrt{B}\int_{\phi _{H}}^{\phi }\frac{d\phi ^{\prime
}}{\sqrt{\Delta \mu (\phi )-\Delta \mu (\phi ^{\prime })}}
\end{eqnarray}
where $\Delta \mu (\phi )$ is given by (\ref{mustar}). Substituting this
integral into (\ref{solAbel}) while noting that $v(0)=\frac{\pi \sigma }{%
2Na^{3}}\phi _{H}$ and $g(z)=2zf(\rho )$, leads to $g(z)$ in the form (\ref%
{gCon}). This equation yields a simple form $z(\phi )$ rather than for $\phi
(z)$. Thus, it is naturally to consider $g(z)$ as a parametric function

\begin{eqnarray}
g(\phi ) &=&\frac{\sigma }{Na^{3}}\sqrt{H^{2}-\Delta \mu (\phi
)/B}\left( \frac{\phi _{H}}{\sqrt{\Delta \mu (\phi
)/B}}+\right.\nonumber
\\&& \left.\sqrt{B}\int_{\phi _{H}}^{\phi }\frac{d\phi ^{\prime
}}{\sqrt{\Delta \mu (\phi )-\Delta \mu (\phi ^{\prime
})}}\right)  \label{g(z)} \\
z(\phi ) &=&\sqrt{\frac{\Delta \mu (\phi _{0})-\Delta \mu (\phi )}{B}}
\end{eqnarray}
These two equations are augmented by (\ref{sig}) relating $\sigma $ to $\phi
_{0}$.

In the case of discontinuous $\phi (z)$ it is necessary to obtain $g(z)$ in
the outer and inner phases separately. For the outer phase the introduction
of the variables $\rho =H^{2}-z^{2}$, $t=H^{2}-z^{\prime 2}$ and $%
g_{-}(z^{\prime })dz^{\prime }=-f_{-}(t)dt$ transforms (\ref{phim}) into an
Abel integral equation (\ref{Abel}) whose solution is (\ref{gm}).

To obtain $g_+(z)$ at the inner phase we substitute (\ref{gm}) into (\ref%
{phip}) and transform the first term of (\ref{phip}) into an Abel integral (%
\ref{Abel}) by introducing the variables $\rho ^{\prime }=H_{t}^{2}-z^{2}$, $%
t^{\prime }=H_{t}^{2}-z^{\prime 2}$ and $g_{+}(z^{\prime })dz^{\prime
}=-f_{+}(t)dt$. This leads to

\begin{equation}
\frac{\pi \sigma }{2Na^{3}}\phi _{+}(z)-\int_{0}^{H^{2}-H_{t}^{2}}\frac{%
f_{-}(t)dt}{\sqrt{\rho -t}}=\int_{0}^{\rho ^{\prime }}\frac{f_{+}(t^{\prime
})dt^{\prime }}{\sqrt{\rho ^{\prime }-t^{\prime }}},
\end{equation}
where $f_{-}(t^{\prime })=g_{-}(z^{\prime })/2z^{\prime }$, and $\rho =\rho
^{\prime }+H^{2}-H_{t}^{2}$. Thus, $v(\rho ^{\prime })$ in the solution of
the Abel equation (\ref{Abel}) is

\begin{equation}
v(\rho ^{\prime })=\frac{\pi \sigma }{2Na^{3}}\phi
_{+}(z)-\int_{0}^{H^{2}-H_{t}^{2}}\frac{f_{-}(t)dt}{\sqrt{\rho -t}}
\label{vrho1}
\end{equation}
and its value at $\rho ^{\prime }=0$ is

\begin{eqnarray}
v(0)&=&\frac{\pi \sigma }{2Na^{3}}\phi _{+}(H_{t})-\int_{0}^{H^{2}-H_{t}^{2}}%
\frac{f_{-}(t)dt}{\sqrt{H^{2}-H_{t}^{2}-t}}\nonumber\\
&=&\frac{\pi
\sigma }{2Na^{3}}\left[ \phi _{+}(H_{t})-\phi _{-}(H_{t})\right] ,
\label{v0}
\end{eqnarray}
where eq. (\ref{Abel}) with $\rho =H^{2}-H_{t}^{2}$ was used in order to
calculate the second term in (\ref{v0}). In addition

\begin{eqnarray}
\frac{dv(\rho ^{\prime })}{d\rho ^{\prime }}&=&\frac{\pi \sigma }{2Na^{3}}%
\frac{d\phi _{+}(z)}{d\rho ^{\prime }}+ \nonumber \\
&&\frac{1}{2}\int_{0}^{H^{2}-H_{t}^{2}} \frac{f_{-}(t)dt}{\left(
H^{2}-H_{t}^{2}+t^{\prime }-t\right) ^{3/2}} \label{dvrho1}
\end{eqnarray}

All together, upon substituting (\ref{vrho1}), (\ref{v0}) and (\ref{dvrho1})
into (\ref{solAbel}) we obtain

\begin{eqnarray}
&&f_{+}(\rho ^{\prime }) =\frac{\sigma }{2Na^{3}}\left[ \frac{\phi
_{+}(H_{t})-\phi
_{-}(H_{t})}{\sqrt{H_{t}^{2}-z^{2}}}+\right.\nonumber
\\
&&\left. \int_{0}^{\rho ^{\prime }}\frac{1}{\sqrt{\rho ^{\prime
}-t^{\prime }}}\frac{d\phi
_{+}(z^{\prime })}{dt^{\prime }}dt^{\prime }\right]  \nonumber \\
&&+\frac{1}{2\pi }\int_{0}^{\rho ^{\prime }}\frac{dt^{\prime }}{\sqrt{\rho
^{\prime }-t^{\prime }}}\int_{0}^{H^{2}-H_{t}^{2}}\frac{f_{-}(t)dt}{\left(
H^{2}-H_{t}^{2}+t^{\prime }-t\right) ^{3/2}}
\end{eqnarray}

In the first integral we express $\rho ^{\prime }$ and $t^{\prime }$ in
terms of (\ref{mustar}) and we change the order of integration in the second
integral obtaining

\begin{eqnarray}
&&f_{+}(\rho ^{\prime }) =\frac{\sigma }{2Na^{3}}\left[ \frac{\phi
_{+}(H_{t})-\phi _{-}(H_{t})}{\sqrt{H_{t}^{2}-z^{2}}}+\right.
\nonumber \\
&&\left.\sqrt{B}\int_{\phi _{+}(H_{t})}^{\phi _{+}}\frac{d\phi
_{+}^{\prime }}{\sqrt{\Delta \mu (\phi
_{+})-\Delta \mu (\phi _{+}^{\prime })}}\right]  \nonumber \\
&&+\frac{1}{2\pi }\int_{0}^{H^{2}-H_{t}^{2}}f_{-}(t)dt \times
\nonumber \\
&&\int_{0}^{\rho ^{\prime }}\frac{dt^{\prime }}{\sqrt{\rho
^{\prime }-t^{\prime }}\left( H^{2}-H_{t}^{2}+t^{\prime }-t\right)
^{3/2}} \label{fp(rho1)}
\end{eqnarray}
The inner integral in (\ref{fp(rho1)}) is

\begin{eqnarray}
&&\int_{0}^{\rho ^{\prime }}\frac{dt^{\prime }}{\sqrt{\rho
^{\prime
}-t^{\prime }}\left( H^{2}-H_{t}^{2}+t^{\prime }-t\right) ^{3/2}}\nonumber \\
&=&\frac{2%
\sqrt{\rho ^{\prime }}}{\left( \rho ^{\prime }+H^{2}-H_{t}^{2}-t\right)
\sqrt{H^{2}-H_{t}^{2}-t}}
\end{eqnarray}
leading to

\begin{eqnarray}
&&f_{+}(\rho ^{\prime }) =\frac{\sigma }{2Na^{3}}\left[ \frac{\phi
_{+}(H_{t})-\phi
_{-}(H_{t})}{\sqrt{H_{t}^{2}-z^{2}}}+\right.\nonumber
\\ &&\left.\sqrt{B}\int_{\phi _{+}(H_{t})}^{\phi _{+}}\frac{d\phi
_{+}^{\prime }}{\sqrt{\Delta \mu (\phi
_{+})-\Delta \mu (\phi _{+}^{\prime })}}\right]  \nonumber \\
&&+\frac{\sqrt{\rho ^{\prime }}}{\pi }\int_{0}^{H^{2}-H_{t}^{2}}\frac{%
f_{-}(t)dt}{\left( \rho ^{\prime }+H^{2}-H_{t}^{2}-t\right) \sqrt{%
H^{2}-H_{t}^{2}-t}}
\end{eqnarray}
Substitution of $f_{-}(t)=g_{-}(z)/2z$ from (\ref{gm}) yields the final
expression for $g_{+}(z)$ in the form

\begin{eqnarray}
&&g_{+}(z)=\frac{z\sigma }{Na^{3}}\left[ \frac{\phi
_{+}(H_{t})-\phi _{-}(H_{t})}{\sqrt{H_{t}^{2}-z^{2}}}+\frac{\phi
_{H}}{\sqrt{H^{2}-z^{2}}}+\right. \nonumber \\
&&\left. \sqrt{B}\int_{\phi _{+}(H_{t})}^{\phi _{+}}\frac{d\phi
_{+}^{\prime }}{\sqrt{ \Delta \mu (\phi _{+})-\Delta \mu (\phi
_{+}^{\prime })}}\right.  \nonumber
\\
&& +\frac{\sqrt{B}}{\pi }\sqrt{H_{t}^{2}-z^{2}} \times
\nonumber \\
&&\int_{0}^{H^{2}-H_{t}^{2}}\frac{dt}{\left( H^{2}-z^{2}-t\right) \sqrt{%
H^{2}-H_{t}^{2}-t}} \times \nonumber \\
&&\left.\int_{\phi _{H}}^{\phi _{\_}^{\prime }(t)}\frac{d\phi
_{-}^{\prime \prime }}{\sqrt{\Delta \mu (\phi _{\_}^{\prime
})-\Delta \mu (\phi _{\_}^{\prime \prime })}}\right] \label{gp}
\end{eqnarray}
It is convenient to express $g_{+}(z)$ in terms of the
concentration of the dense phase, $\phi _{+}$ utilizing $z(\phi
_{+})\-=\sqrt{H^{2}-\Delta \mu (\phi _{+})/B}$. Introducing the
variables $t\-=\Delta \mu (\phi _{\_}^{\prime })/B$ and
$dt\-=1/B(d\Delta \mu (\phi _{\_}^{\prime })/d\phi _{\_}^{\prime
})d\phi _{\_}^{\prime }$ in the last integral leads to
(\ref{gp(z)}).

\section{Model concentration profiles\label{appendixModel}}

A rough approximation allows to obtain analytical expression for $g(z)$ of a
brush in the presence of a vertical phase separation. In the dense phase the
variation of $\phi $ is slow and we can approximate it as constant, $\phi
_{+}=\phi _{0}$. In the outer phase dilute phase $\phi $ is rather low and
we can neglect nonconstant terms in $\overline{\chi }(\phi )$. Thus

\begin{equation}
\phi (z)=\left\{
\begin{array}{c}
\phi _{0},0<z<H_{t} \\
\phi (z),H_{t}<z<H%
\end{array}
\right.
\end{equation}
where $\phi (z)$ is determined by (\ref{mustar}) with $\chi =const$, while
the value of $\phi _{0}$ is set by(\ref{ConfN1}). As before, $g(z)$ in the
outer phase is determined by (\ref{gm}) while in the inner phase it is
determined by (\ref{gp}).

First, consider the case of $\chi =1/2$ leading to $\Delta \mu (\phi )=-\ln
(1-\phi )-\phi \approx \phi ^{2}/2$ and eq. (\ref{gm}) yields

\begin{equation}
g(z)=z\frac{\sigma \pi }{2Na^{3}}\sqrt{2B}  \label{gmm}
\end{equation}

Substitution of (\ref{gmm}) into (\ref{gp}) gives the expression for $g(z)$
in the inner phase

\begin{eqnarray}
g(z)&=&\frac{z\sigma }{Na^{3}}\left[ \frac{\phi _{+}(H_{t})-\phi _{-}(H_{t})}{%
\sqrt{H_{t}^{2}-z^{2}}}+\right.\nonumber \\
&&\left.\sqrt{2B}\arctan \sqrt{\frac{H^{2}-H_{t}^{2}}{%
H_{t}^{2}-z^{2}}}\right]  \label{gpm}
\end{eqnarray}
where $\phi _{-}(H_{t})=\sqrt{2B(H^{2}-H_{t}^{2})}$ and $\phi
_{+}(H_{t})=\phi _{0}$. This function is discontinuous and diverges at the
phase boundary $z=H_{t}$. The value of $\phi _{0}$ can be found from (\ref%
{ConfN1})

\begin{equation}
\phi _{0}=\frac{Na^{3}}{\sigma H_{t}}-\sqrt{\frac{B}{2}}\left[ \frac{H^{2}}{
H_{t}}\left( \frac{\pi }{2}-\arcsin \frac{H_{t}}{H}\right) -\sqrt{
H^{2}-H_{t}^{2}}\right]
\end{equation}

When $\chi =0$ eq. (\ref{gm}) specifies $g(z)$ at the outer phase, $%
H_{t}<z<H $

\begin{equation}
g(z)=z\frac{\sigma }{Na^{3}}B\sqrt{H^{2}-z^{2}}.
\end{equation}
and eq. (\ref{gp}) for $g(z)$ in the inner phase, $0<z<H_{t}$, yields

\begin{figure}
\begin{center}
\includegraphics[width=7cm]{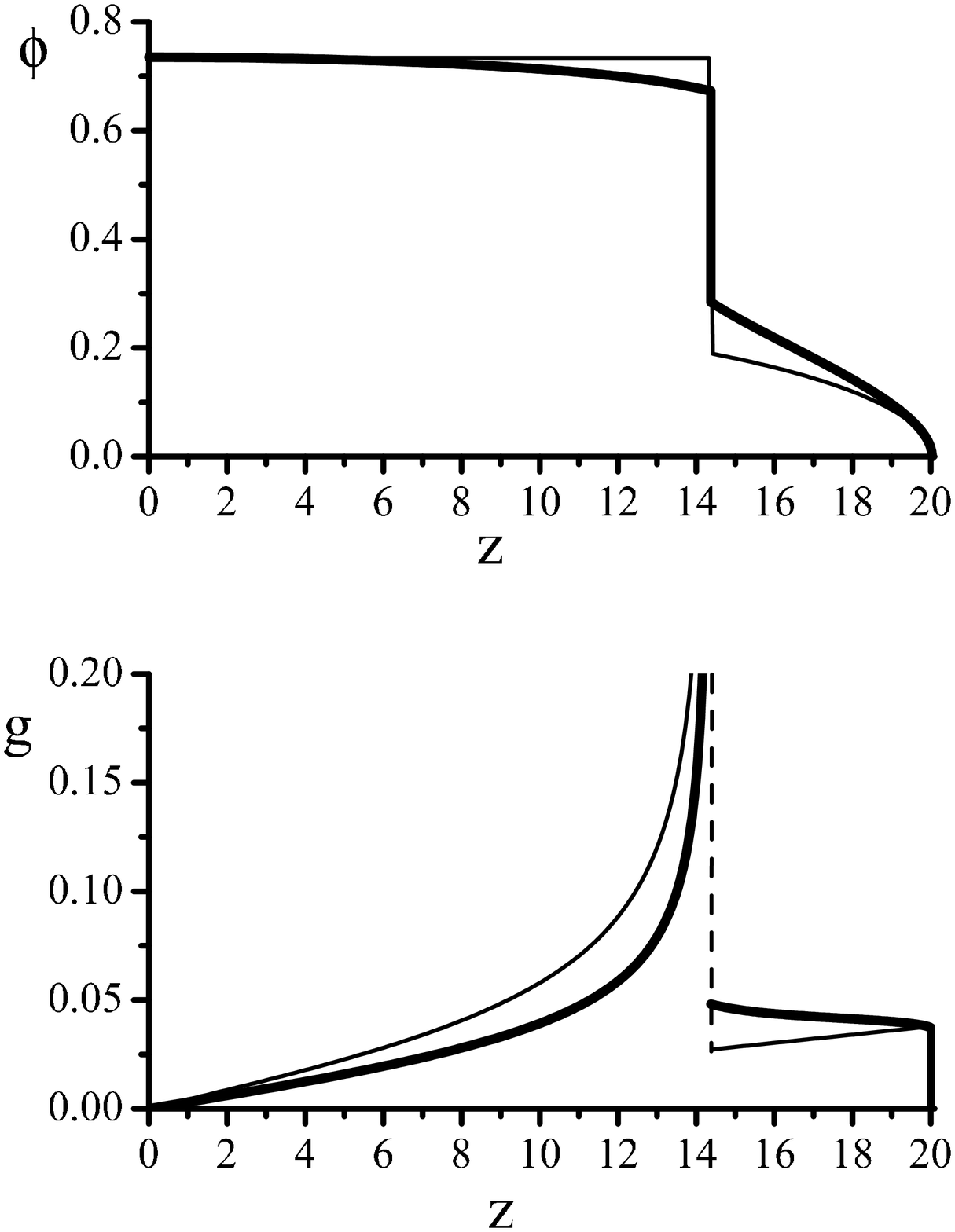}
\end{center}
\caption{Comparison between the exact $\phi (z)$ and $%
g(z)$ and their approximate values as calculated from for the case of $%
\overline{\chi}(\phi )=1/2+1.05\phi ^{2}$, $\sigma =120$ and $N=300$.\cite%
{FootFig}} \label{fig13}
\end{figure}

\begin{eqnarray}
g(z)&=&\frac{z\sigma }{Na^{3}}\left[ \frac{\phi _{+}(H_{t})-\phi _{-}(H_{t})}{%
\sqrt{H_{t}^{2}-z^{2}}}+\right.\nonumber \\
&&\left. 2B\left( 2\sqrt{H^{2}-H_{t}^{2}}+\sqrt{%
H_{t}^{2}-z^{2}}-\sqrt{H^{2}-z^{2}}\right) \right]
\end{eqnarray}
Again, $g(z)$ of the inner phase diverges at the phase boundary. The
performance of this approximation is illustrated in Fig \ref{fig13}. It
captures the main features of $\phi (z)$ and the behavior of $g(z)$ in the
inner region. However $g(z)$ at the outer region increases rather than
decrease.

\pagebreak


\begin{thebibliography}{99}
\bibitem{foot2} This expression can be inverted as $\chi _{eff}(\phi )=\-\frac{%
\chi _{eff}(0)-\int_{0}^{\phi }\overline{\chi }(\phi )d\phi
}{1-\phi }$.
\bibitem{FootFig} All calculation utlize equation (\ref{sig}) or (\ref%
{sigdis}) and begin with choosing $\phi_{0}$. This is then used to determine
$H$ and then $\sigma $. For numerical convenience the calculation of $\sigma
$ is carried out by using discretized version of equation (\ref{sig}) or (%
\ref{sigdis}). The integral is approximated by a sum of $100$ equidistant
terms. It is thus difficult to control $\sigma $ precisely. The $\sigma $
stated values are the integr values within $.2$ of the real value.
\end{thebibliography}
\end{document}